\documentclass[notitlepage,twocolumn,prl,tightenlines,nofootinbib,superscriptaddress,showpacs]{revtex4}

\usepackage{amsmath,amssymb,amsfonts,latexsym}
\usepackage{bbm}
\usepackage[mathcal]{euscript}
\usepackage{graphicx}
\usepackage{epsfig}
\usepackage{color}
\usepackage[activate=normal]{pdfcprot}  

\usepackage{psfrag}
\usepackage{txfonts,pxfonts,wasysym,}
\usepackage{pifont}
\usepackage[caption = false]{subfig}

\newcommand{\gulli}{UMR Gulliver 7083 CNRS, ESPCI ParisTech,
PSL Research University, 10 rue Vauquelin, 75005 Paris, France}
\newcommand{\uqroo}{CONACYT - Universidad de Quintana Roo, Boulevard Bah\'ia s/n, Chetumal, 77019 Quintana Roo, M\'exico}

\newcommand{\be}{\begin{equation}}
\newcommand{\ee}{\end{equation}}
\newcommand{\ben}{\begin{equation*}}
\newcommand{\een}{\end{equation*}}
\newcommand{\ba}{\begin{eqnarray}}
\newcommand{\ea}{\end{eqnarray}}

\begin{document}
\graphicspath{{./figures/}}

\title{Active vs. Passive Hard Disks Against a Membrane : Mechanical Pressure and Instability}

\author{G. Junot}
\affiliation{\gulli}
\author{G. Briand}
\affiliation{\gulli}
\author{R. Ledesma-Alonso}
\affiliation{\gulli}
\affiliation{\uqroo}
\author{O. Dauchot}
\affiliation{\gulli}

\date{\today}

\begin{abstract}
We experimentally study the mechanical pressure exerted by a set of respectively passive isotropic and self-propelled polar disks onto two different flexible unidimensional membranes. In the case of the isotropic disks, the mechanical pressure, inferred from the shape of the membrane, is identical for both membranes and follows the equilibrium equation of state for hard disks. On the contrary, for the self-propelled disks, the mechanical pressure strongly depends on the membrane in use, and is thus not a state variable. When self propelled disks are present on both sides of the membrane, we observe an instability of the membrane akin to the one predicted theoretically for Active Brownian Particles against a soft wall. In that case, the integrated mechanical pressure difference across the membrane can not be computed from the sole knowledge of the packing fractions on both sides; a further evidence of the absence of equation of state.
\end{abstract}

\maketitle
Developing a thermodynamics for out of equilibrium systems of dissipative particles, to which energy is homogeneously provided at the particle level, has been a long lasting effort driven by the impressive success of equilibrium thermodynamics. An intense activity dealt with the search for an effective temperature and a generalized Gibbs measure in granular media~\cite{Coniglio01a,Makse2002a,blumenfeld2003gee,visco2005ipa,Lechenault:2006id,Bertin:2006jm,Song:2008jb,AbatePRL08,Behringer:2014cd}. More recently, defining pressure in systems of active particles, the specificity of which is to turn the injected energy into directed motion, has attracted a lot of attention~\cite{Yang:2014bl,Mallory:2014dla,Takatori:2014do,Takatori:2015ic,Yan:2015fg,Yan:2015cf,Solon:2015hza,Solon:2015bt,Winkler:2015cy,Hancock:2017vp}. In both cases, the intensity of the scientific debates highlights the importance of the addressed issues.

At equilibrium, the mechanical, hydrodynamic and thermodynamic pressures are all equal quantities, which inherit their mutual properties. In particular equilibrium pressure is a state variable, which only depends on the bulk properties of the fluid. 
On the theoretical side, the fact that some form of pressure could be considered as a state variable, hence obeying an equation of state (EOS), has been the topic of intense debates in a series of recent works~\cite{Yang:2014bl,Mallory:2014dla,Takatori:2014do,Takatori:2015ic,Yan:2015fg,Yan:2015cf,Solon:2015hza,Solon:2015bt,Winkler:2015cy,Hancock:2017vp}. More specifically, probing the mechanical pressure exerted on a wall, it was shown that the existence of an EOS strongly depends on the details of the dynamics and in general does not hold~\cite{Solon:2015bt}. Different conclusions were obtained when considering the virial pressure defined in the bulk of an active suspension~\cite{Takatori:2014do,Takatori:2015ic,Yan:2015fg}. Wether these debated results hold for realistic systems remains anyhow unclear.
On the experimental side, we know of only one result which infers the osmotic pressure of sedimenting active particles from the measurement of their density profiles~\cite{Ginot:2014vk}. To our knowledge, no direct measurement of mechanical quantities have been reported.\\
%
\begin{figure}[t!] 
\includegraphics[width=0.45\columnwidth,trim=0mm 0mm 0mm 0mm,clip]{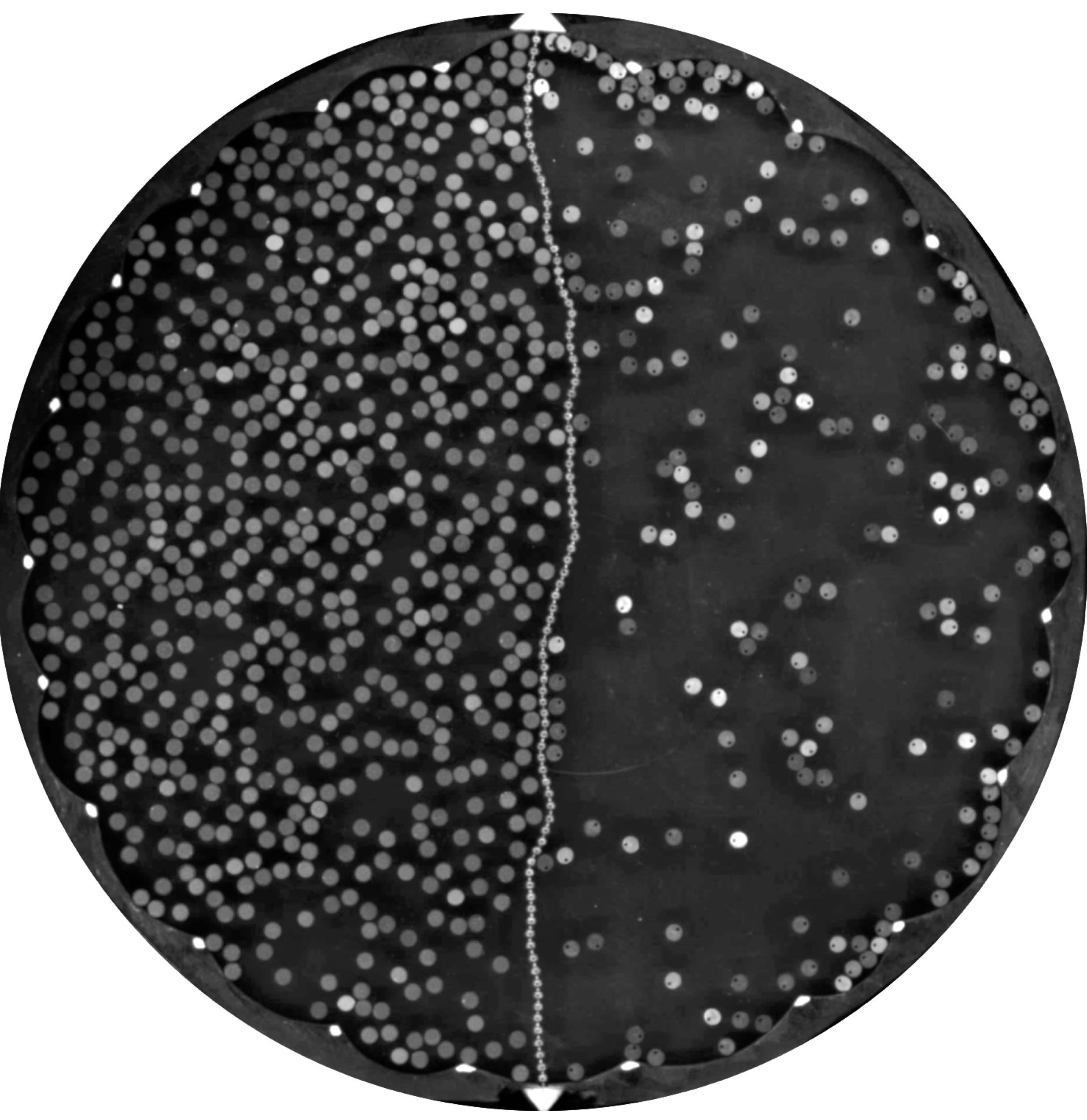}
\hspace{2mm}
\includegraphics[width=0.45\columnwidth,trim=0mm 0mm 0mm 0mm,clip]{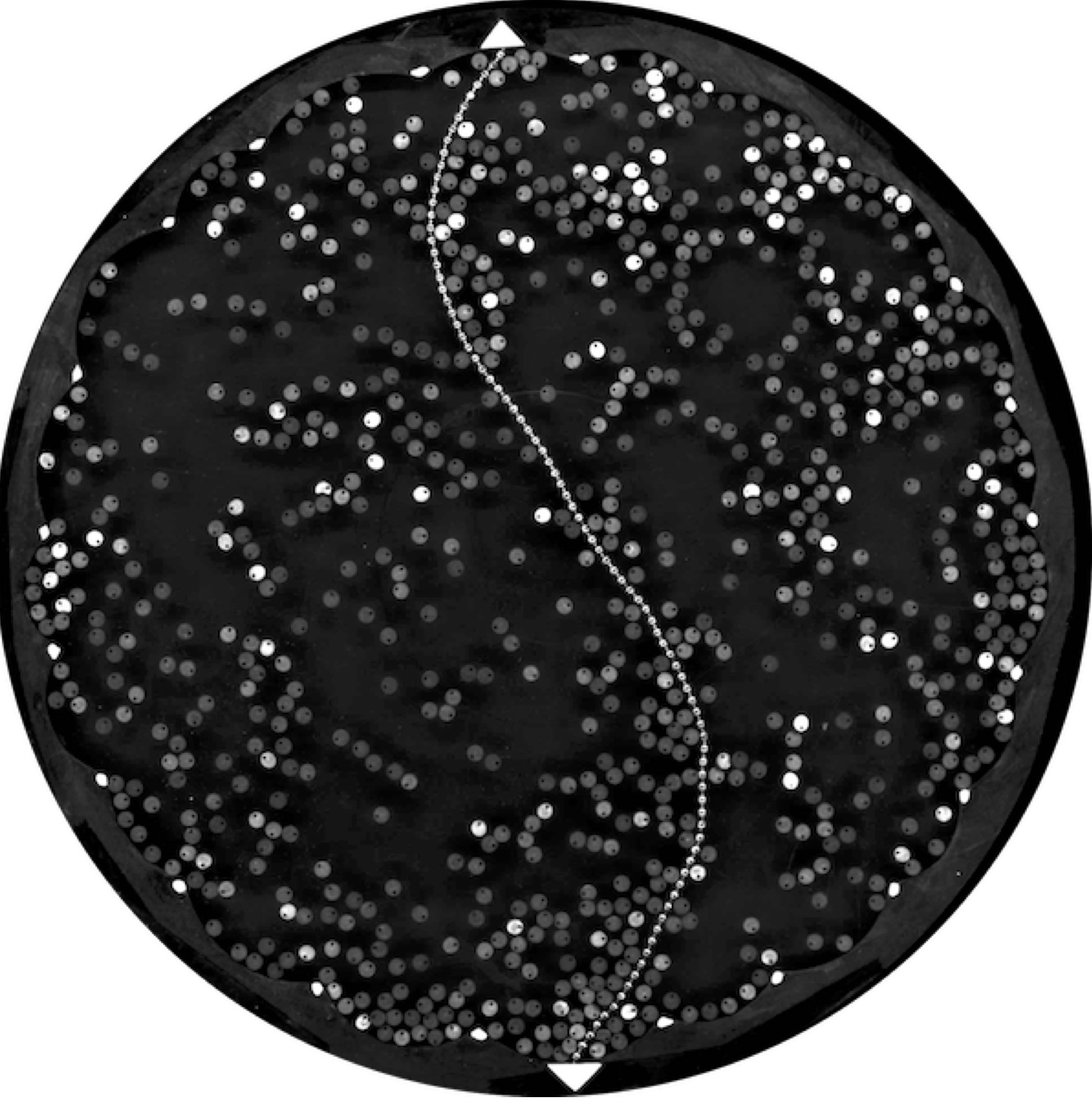}
\hbox{\hspace{-0.4\columnwidth} (a) \hspace{0.45\columnwidth} (b)}
\vspace{-0mm}
\caption{ {\bf Mechanical Equilibrium and Instability.} Self-propelled polar disks and/or isotropic passive disks are distributed on the two sides of a vibrating chain attached at both ends, acting as a separating membrane: (a)  mechanical equilibrium between a small number of self propelled disks (on the right) and a large number of isotropic passive disks (on the left); (b) S-shape instability observed when self-propelled disks are distributed in equal quantities on both sides of the membrane.}
\label{fig:instability}
\vspace{-0.5cm}
\end{figure}
In this letter, we take advantage of a 2D experimental system of vibrated disks~\cite{Deseigne:2010gc,Deseigne:2012kn}, to investigate the mechanical pressure exerted by assemblies of respectively passive and active disks on two membranes made of vibrated chains. Our main results are as follow: (i) when either self propelled polar or passive isotropic disks are introduced on one side of the arena, the tension in the membrane equilibrates with an homogeneous mechanical pressure, which obeys the Laplace's law; (ii) quite remarkably, in the case of the isotropic disks, the pressure follows the equilibrium equation of state for hard disks; (iii) in the case of the self-propelled disks, the pressure depends on the selected chain, emphasizing that mechanical pressure is not a state variable; (iv) for a given chain the mechanical equilibrium between self propelled and isotropic disks [Fig.\ref{fig:instability}(a)] is still set by the equality of the mechanical pressure; (v) when introducing self propelled disks on both sides, an instability of the membrane [Fig.\ref{fig:instability}(b)], akin to the one predicted theoretically for Active Brownian Particles (ABP) against a flexible wall~\cite{Nikola:2016jca}, takes place. Altogether our results demonstrate that setting up a thermodynamical frame for active particles remains an important conceptual challenge, but also offers new design opportunities in adapting wall properties to particles specificities.

The experimental system is made of vibrated disks with a built-in polar asymmetry, which enables them to move coherently~\cite{Deseigne:2012kn}. The polar particles are micro-machined copper-beryllium disks (diameter $d = 4$ mm, area $a = \pi d^2/4$) with an off-center tip and a glued rubber skate located at diametrically opposite positions (total height $h = 2$ mm).  These two ``legs'', with different mechanical response, endow the particles with a polar axis. Under proper vibration, the self propelled polar (spp) disks perform a persistent random walk, the persistence length of which is set by the vibration parameters. Here we use a sinusoidal vibration of frequency $f=95$ Hz and amplitude $A$, with relative acceleration to gravity $\Gamma = 2\pi A f^2/g = 2.4$. We also use plain rotationally-invariant disks (same metal, diameter, and height), hereafter called the ``isotropic'' (iso) disks. 
A chain -- the type of chain used to attach bathtub drain stoppers -- is formed of $J+1$ beads of diameter $\sigma$, connected by $J$ rigid links, with a center to center distance between two beads $\ell$. The joint between a bead and a link is torque free, but the angle between successive links can not exceed $\eta_{max}$ radians. The chain is fixed at its ends, dividing the arena diametrically. The total length of the chain $L_0 = J\ell$ is longer than the end-to-end distance of the chain, the arena diameter $D=251.1$ mm. In the following we use two chains, the geometrical properties of which are summarized in table 1.
The instantaneous position of the beads along the chain is captured using a CCD camera at a frame rate of $1$ Hz during $60$ min, after an equilibration time of $30$ min. In the following, the unit of time is set to be the inverse frame rate and the unit length is the particle diameter $d$. 
\begin{table}[b]
\vspace{-5mm}
\begin{center}
\begin{tabular}{|l|c|c|c|c|c|}
		& J     & $\sigma$ (mm)  & $\ell$ (mm)  & $\eta_{max} $ & $L_0$  (mm) \\
\hline
chain-1 	& 91   & 	2.33 		   & 	3.10 	  & 	$\pi/8$ 	    	& 282.10 		\\
chain-2 	& 147 &    1.44 		   & 	1.92   & 	$\pi/8$ 		& 282.24 		\\ 
 
\end{tabular}
\caption{Geometric parameters of the chains}
\end{center}
\label{table:chain}
\vspace{-4mm}
\end{table}
%
\begin{figure}[t!]
\centering
\includegraphics[width=0.95\columnwidth]{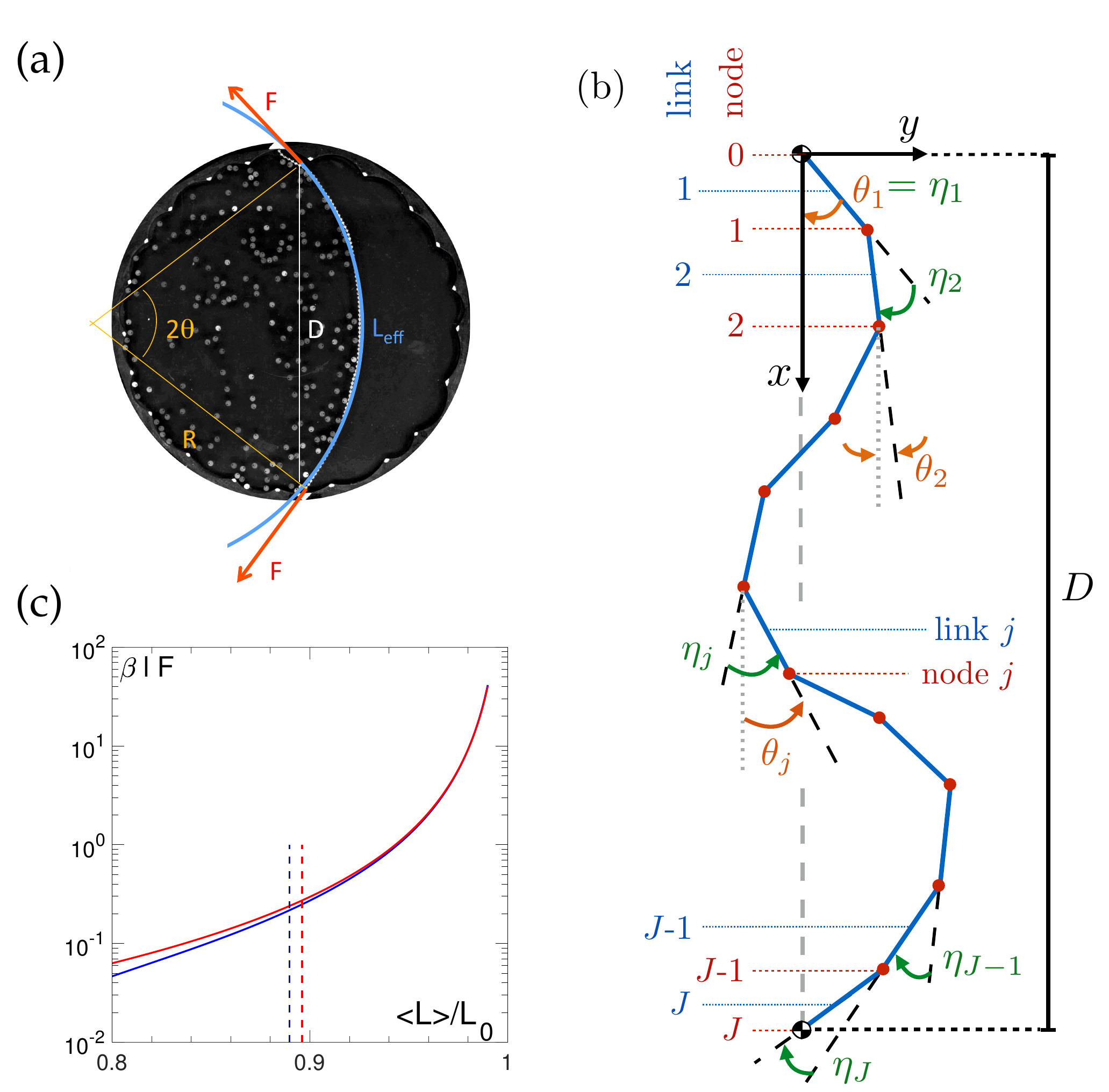}
\vspace{-3mm}
\caption{{\bf Membrane barometer.} (a) Instantaneous configurations of chain-1 separating the system into two compartment, one filled with $640$ isotropic particles, and the second left empty. The blue curve is the \emph{effective membrane} of length $L_{eff}$, which takes the form of an arc-circle of radius $R$.  (b) Sketch of the simplified chain model: rigid links are connected by torque free joints. The angle between two successive links is bounded to $\eta_{max}$. (c) Effective elasticity : tension $\mathcal{F}=\beta\ell F$ in a model chain as a function of its normalized averaged length $\left<L\right>/L_0$ for chain-1 (blue) and chain-2 (red); the vertical dotted lines indicate $D/L_0$}
\label{fig:setup}
\vspace{-0.5cm}
\end{figure}

The chain vibrates and explores all configurations compatible with the constraint on the angles between successive links $\eta_j \leq \eta_{max}$. The positions of the successive beads, averaged over $3600$ samples, separated by $5700$ vibration cycles, define a line of length $L_{eff}$, which we shall call \emph{the (effective) membrane}. In the absence of particles, it aligns along the diameter of the arena and $L_{eff} = D$.
As soon as a few hundred particles, whether isotropic or polar, are introduced on one side of the chain, the symmetry is broken. Apart from a boundary layer of one or two links at the extremities, the membrane takes the shape of an arc-circle, with constant radius of curvature $R$ and length $L_{eff} \in \left[D\,\,  L_0\right]$. The angles between the successive links composing the membrane are all smaller than $\eta_{max}$ [Fig.~\ref{fig:setup}(a)]. Hence the membrane does not support any torque; the tension $F$ in the membrane is constant; and the mechanical pressure $\Pi$ is homogeneous. Mechanical equilibrium then leads to:
\vspace{-1mm}
\be
\Pi \times D = 2 F \sin\left(\frac{L_{eff}}{2R}\right)\ ,
\ee
Together with the purely geometric relation:
\be
\frac{D}{2R} = \sin\left(\frac{L_{eff}}{2R}\right)\ ,
\ee
\vspace{-1mm}
one finds that the the mechanical pressure obeys the Laplace's law, $\Pi=F/R$, which describes the pressure difference across an interface between two fluids~\cite{deGenne2003cwp}. 

To proceed further, we compute the effective elasticity of the membrane, which relates its effective length $L_{eff}$ to its tension $F$,  \emph{assuming} that it obeys an equilibrium dynamics. To do so, we consider a model chain composed of J beads, and compute its average end to end distance $\left<L\right>$ under an imposed tension $F$. We proceed in two steps. First, one extremity being fixed, the other being free of force, we evaluate the probability density of its end to end distance $L$. The angles $\eta_j$ are generated randomly within the range $\left[-\eta_{max},\eta_{max}\right]$, from a uniform distribution with bins $\delta\eta=10^{-3}$ rad. The distribution $\rho\left(L,J\right)$ is then obtained from $5.10^9$ configurations using :

\begin{equation}
L=\ell\left\{\left[\sum_{j=1}^J\cos\left(\theta_j\right)\right]^2+\left[\sum_{j=1}^J\sin\left(\theta_j\right)\right]^2\right\}^{1/2} \ ,
\end{equation}
where $\displaystyle\theta_j=\sum_{i=1}^j \eta_i$, for $j=1,\dots ,J$ [Fig.~\ref{fig:setup}(b)].
Second, the partition function for a chain with end-to-end distance $L$ in the presence of a tension $F$ is given by:
\begin{equation}
\dfrac{Z\left(\beta \ell F,J\right)}{Z_0\left(J\right)}=\int_{0}^{L_0}\rho\left(L, J\right)\exp\left(\beta L F \right) d L \ ,
\end{equation}
where $Z_0=\left(2\eta_{max}\right)^J$ is the total number of configurations in the limit $\beta\rightarrow 0$, and $\beta$ is the equilibrium temperature of the chain.
The above chain model is very similar to the 2D version~\cite{Jagota2014poly} of the well-know ``freely jointed chain"~\cite{Mazars1996fjc}, the main difference being the maximum angle constraint between links.
Our model thus shows an end-to-end length distribution of intermediate type between the 2D freely jointed chain~\cite{Jagota2014poly} and the worm-like chain model~\cite{Frey1996poly}.

Finally, the average length of the chain $\langle L\rangle$ reads~\cite{Rubinstein2003pph} :
\vspace{-2mm}
\begin{equation}
\langle L\rangle(F,J) = \dfrac{1}{\beta} \left[\dfrac{\partial}{\partial F}\ln\left(Z\right)\right]_{J}  \ .
\end{equation}
\vspace{-1mm}
Figure~\ref{fig:setup}(c) displays the result of the above analysis in the form $\beta \ell F=\mathcal{F}\left(\dfrac{\langle L\rangle}{L_0}\right)$, which we shall interpret as the effective elasticity of the membranes. The two membranes obey almost identical effective elasticity in the range of length $D<L_{eff}<L_0$. Note that they however differ by their dynamical properties: the relaxation time $\tau$ -- evaluated as the time to relax a typical shape fluctuation -- is significantly shorter for chain-1 ($\tau_1=15$), than for chain-2 ($\tau_2=90$). 

We are now in position, measuring the radius of curvature $R$ and the length $L_{eff}$ of the membrane to access the dimensionless mechanical pressure:
\begin{equation}
\Pi^{\ast} = \beta \Pi a  =  \dfrac{a}{\ell R}\mathcal{F}\left(\dfrac{L_{eff}}{L_0}\right).
\label{Eq:PFunc}
\end{equation}
%
\begin{figure}[t]
\centering
\includegraphics[width=0.38\columnwidth,trim=15mm 27mm 35mm 35mm,clip]{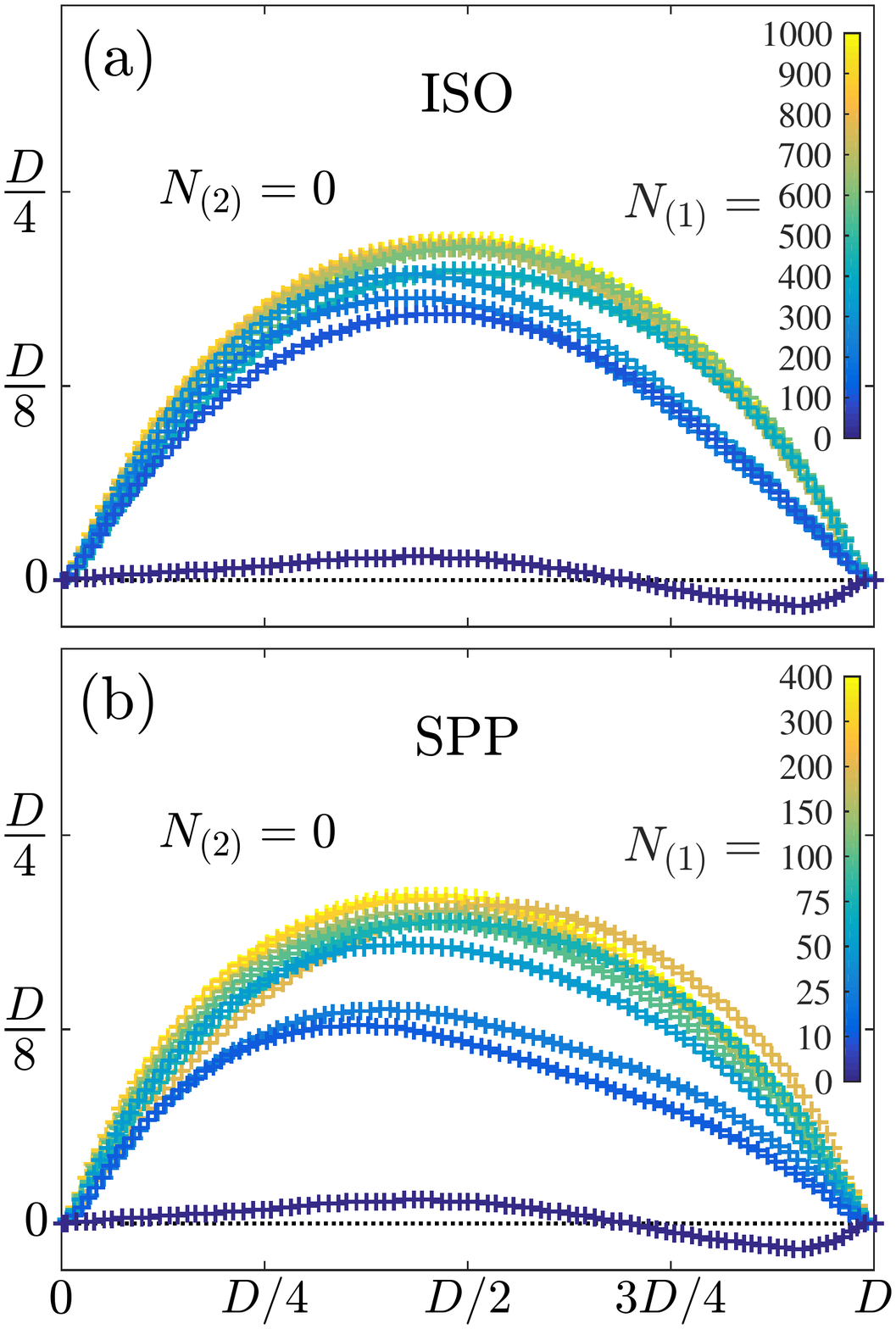}
\hspace*{0mm}
\includegraphics[width=0.60\columnwidth,trim=0mm 5mm 0mm 0mm,clip]{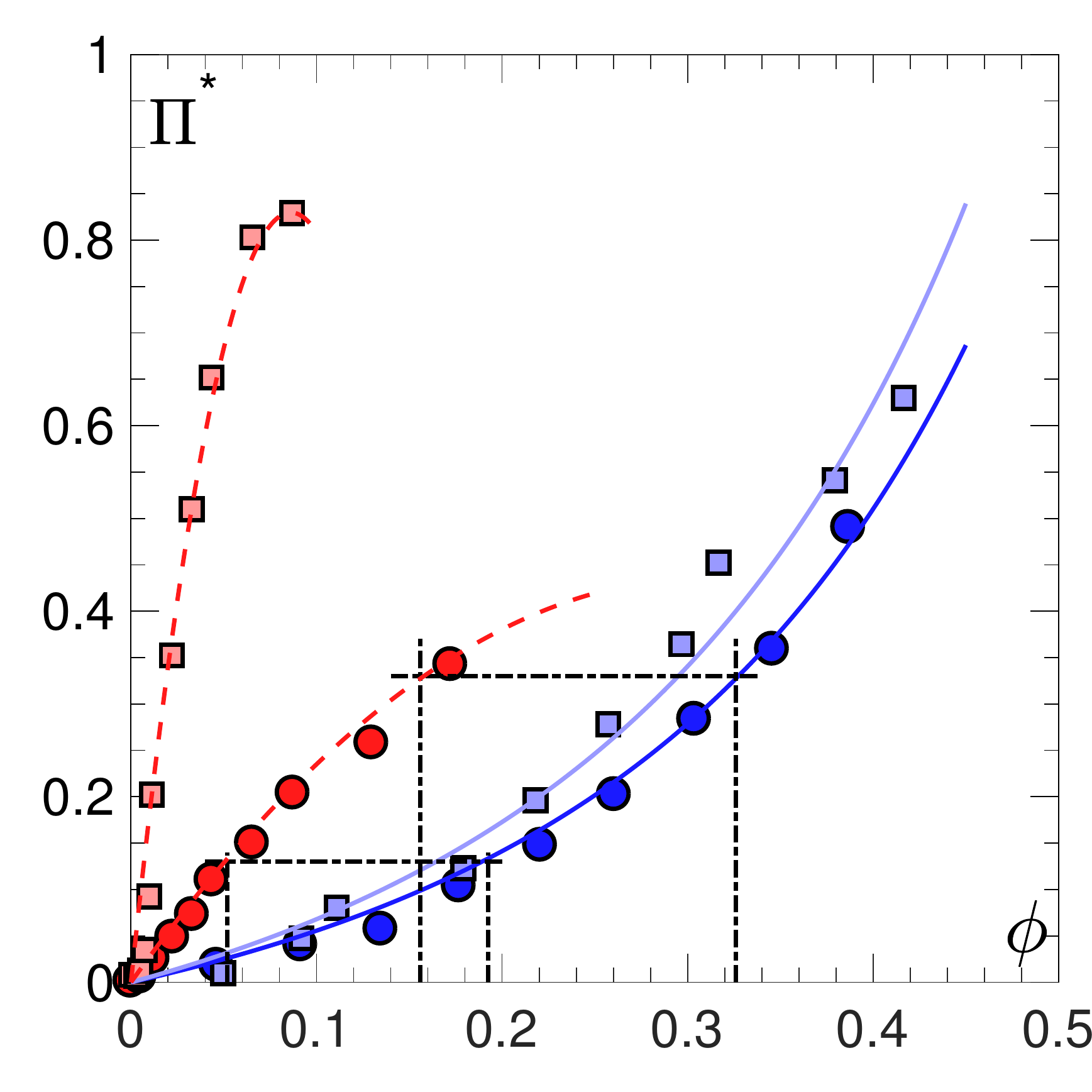}
\caption{{\bf Pressure.} Left : average chain-1 configurations when disks are confined to a single region, with (a) $N_1=[0,100,200,...900]$ isotropic disks (iso) and (b) $N_1=[0, 10, 25, 50, 75, 100, 150, 200, 300, 400]$ self propelled polar disks (spp).
Right : dimensionless pressure $\Pi^{\ast}$ as a function of the packing fraction $\phi$, for the isotropic (blue) and polar (red) disks, for chain-1 ($\circ$) and chain-2 ($\square$); the plain curves are the equilibrium equation of state for hard disks (eq.\eqref{Eq:ZCS}) -- at different temperatures; the dashed curves are obtained from a $2^{nd}$ order virial expansion (eq.\eqref{Eq:Virial}). The dot-dashed lines point at the configurations for which mechanical equilibrium between spp and iso disks have been probed.}
\label{fig:pressure}
\vspace{-0.3cm}
\end{figure}
The experimental results are summarized in Fig.~\ref{fig:pressure}. For both the isotropic and the polar disks, the chain indeed takes the shape of an arc-circle when the number of particles is large enough, say larger than a  hundred. For less particles, we observe some distorsion, which in the case of the isotropic disks can reasonably be attributed to statistical fluctuations. In the case of the polar particles, they are more significant and we shall come back to it below. 
For the moment, we assume a perfect arc circle shape, extract its curvature radius and compute the mechanical pressure using eq.~\eqref{Eq:PFunc}. 

In the case of the isotropic disks, $\Pi^{\ast}(\phi)$, is well described by the thermodynamic equation of state $P_{HD}(\phi)$ for hard disks~\cite{Santos:1995hz}, up to a multiplicative constant :
\vspace{-1mm}
\begin{equation}
P^{\ast}_{HD} = \beta P_{HD} a = \frac{\beta}{\beta_{HD}} \phi\ \frac{1+\alpha\phi^2}{\left[1-\phi\right]^2} \ ,
\label{Eq:ZCS}
\end{equation}
\vspace{-0.5mm}
where $\alpha = 7/3 - 4\sqrt{3}/\pi$. The multiplicative constant $\beta/\beta_{HD}$ accounts for the fact that the effective temperatures of the chains and that of the disks have no reason to be equal ($\beta/\beta_{HD}=0.45$ for chain-1 and $\beta/\beta_{HD}=0.55$ for chain-2). As a matter of fact, it is already very remarkable that (i) the equilibrium assumption used to describe the membranes holds and (ii) that the isotropic disks can also be described  
within an equilibrium framework. 

In the case of the self-propelled polar disks, we observe a very different dependance of the mechanical pressure $\Pi^{\ast}$ on the packing fraction with a clear concave shape. The data can be captured by a Virial expansion at the second order in $\phi$:
\begin{equation}
\Pi^{\ast}_{spp} =  \beta \Pi_{spp} a = \frac{\beta}{\beta_{spp}} \phi \left(1+b_1\phi\right)+O\left(\phi^3\right)
\label{Eq:Virial}
\end{equation}
with a negative $b_1$, which accounts for the concave shape. We find $\beta/\beta_{spp} = [6, 35]$ and   $b_1=[-1.6, -5.8]$, for respectively chain-1 and chain-2. The mechanical pressure $\Pi^*(\phi)$ for the self propelled particles, when measured with the two different chains, follows different dependences on the packing fraction, which cannot be absorbed in a multiplicative factor : there is no equation of state.

We now consider the equilibration of the mechanical pressure induced by $N_{spp}$ self propelled disks on one side of the chain by $N_{iso}$ isotropic disks on the other side [Fig~\ref{fig:instability}(a)]. The experiment is first conducted with chain-1 and for $N_{spp}=100, 300$. As demonstrated from the dot-dashed lines on fig.~\ref{fig:pressure}(c), the number of isotropic disks $N_{iso}=370, 627$ required to equilibrate the mechanical pressure imposed by the self propelled disks is prescribed by the equality of the mechanical pressure $\Pi^{\ast}_{iso}(\phi_{iso})=\Pi^{\ast}_{spp}(\phi_{spp})$, which has been measured independently. When mechanical equilibrium holds for chain-1 (say with $N_{spp} = 300$ and $N_{iso} = 627$), and one replaces chain-1, by chain-2, the equilibrium is broken, the spp developing a much larger pressure. It is restored for only $N_{spp }= 45$, further stressing the dependance of the mechanical equilibrium on the chain properties.

\begin{figure}[t!]
\centering
\includegraphics[width=0.95\columnwidth]{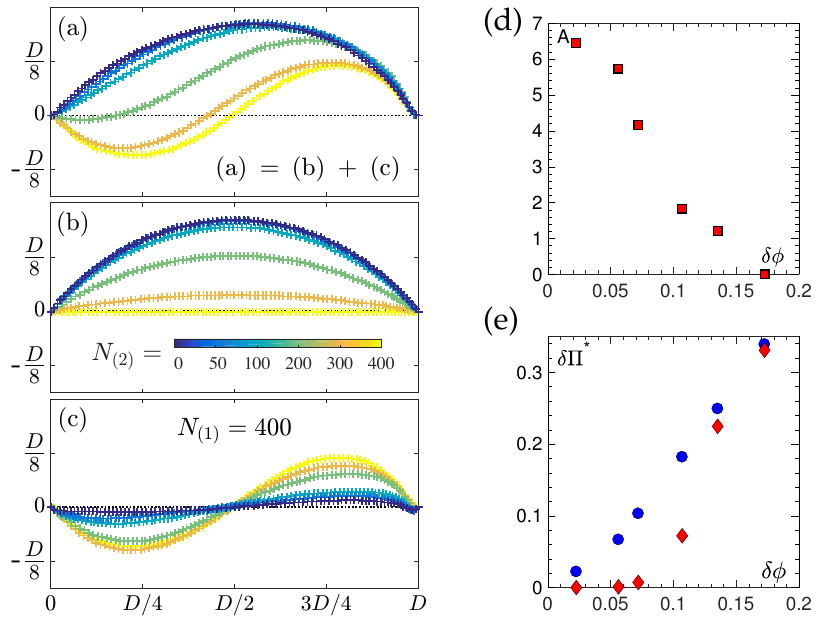}
\vspace{-2mm}
\caption{{\bf Mechanical instability.} (a) Average chain configuration for spp, which are distributed in the two regions, $N_{(1)}=400$ and several values of $N_{(2)}\leq N_{(1)}$. The chain configuration (a) decompose into (b) a circular segment plus (c) a sinusoidal shape. (d) Amplitude of the sinusoidal modulation $A$ as a function of the packing fraction difference $\delta\phi$. (e) Integrated pressure difference as a function of the packing fraction difference $\delta\phi$, obtained ($\diamond$) from the geometry of the average chain and eq.\eqref{Eq:PFunc} and ($\circ$) from the independent measurement of $\phi_1$ and $\phi_2$, and the virial expression~(\ref{Eq:Virial}).}
\label{fig:decompo}
\vspace{-0.5cm}
\end{figure}

Finally, we consider the situation where the two containers are filled with self-propelled disks. Instead of displaying the now familiar arc circle configuration, the membrane takes a series of amazing shapes with inhomogeneous curvature. The most spectacular case takes place when there is an identical number of disks on both sides~: instead of fluctuating around an average flat configuration, the membrane takes a well pronounced S-shape [Fig.~\ref{fig:instability}(b)], which is very much reminiscent of the one adopted by a flexible filament with fixed ends immersed in an active gas of Active Brownian Particles~\cite{Nikola:2016jca}.  
While the mechanical pressure is certainly not uniform along the chain, the symmetry of the S-shape pattern suggest that the integrated pressures along the chain are equal on both sides. To gain further insight in that matter, we systematically vary the number of polar disks on both sides with $N_1 = [100, 200, 300, 400]$ and $N_2 < N_1$. 
A remarkable fact is that for all pairs $(N_1,N_2)$, the a priori complicated shape of the average chain can be decomposed into an arc-circle plus a modulation of wavelength $D$, as illustrated on figure~\ref{fig:decompo}(a-c) for the case $N_1= 400$, $N_2=[0,100,200,300,400]$. As shown on figure~\ref{fig:decompo}(d), the amplitude of the sinusoidal modulation is maximal when $N_1=N_2$ (packing fraction difference close to zero, $\delta\phi\rightarrow0$) and approaches zero when one side is empty. The above systematic deviation of the chain shape from a perfect arc-circle, alluded to in the previous section, thus finds its origin in the present instability. 
At first order, the sinusoidal part being odd does not contribute to the integrated pressure difference between the two sides, $\delta\Pi^{\star}$.  We thus extract the latter using the same procedure as above and compare it to $\Pi_{spp}^{\star}(\phi_1) - \Pi_{spp}^{\star}(\phi_2)$, obtained from the measurement of the packing fractions $\phi_1$ and $\phi_2$, and the virial expression~(\ref{Eq:Virial}). In the limiting cases, where either one side is empty or where both sides are equally filled in number of particles, both should be equal by construction. It is however clear from figure~\ref{fig:decompo}(e) that the equality only holds in these two limits. In between, when there is an unequal number of self-propelled particles on both sides, the mechanical pressure across the membrane can not be computed from the sole knowledge of the packing fractions on both sides. This further demonstrates the absence of EOS.
%

Altogether, while the mechanical pressure exerted by a gas of shaken isotropic granular disks against a membrane essentially obeys the EOS for equilibrium hard disks -- a non trivial result --, the one exerted by a gas of self propelled polar disks obeys a very different physics.
First, it increases much faster with the packing fraction, and present a negative value of the first Virial coefficient. A similar observation was made for the osmotic pressure measured in a suspension of self-propelled phoretic Janus colloids~\cite{Ginot:2014vk}. 
Second, it depends on the properties of the membrane itself and is thus not a state variable. Our results reinforce those theoretical analysis, which conclude to the generic absence of EOS for active particles~\cite{Solon:2015bt,Nikola:2016jca}. They may however look surprising in regards of the fact that Active Brownian disks do obey an EOS~\cite{Yang:2014bl,Mallory:2014dla,Takatori:2014do,Yan:2015cf,Solon:2015hza}. The reason is that this last result only holds when the propulsive force decouples from the orientational dynamics~\cite{Solon:2015bt}. Here it is not the case: the orientation dynamics obeys a torque coupled to the velocity dynamics~\cite{Weber:2013bj}. It would be interesting to perform a perturbative analysis of the effect of such coupling around the ideal ABP case.
Finally, the membranes being essentially identical from the point of view of their effective elasticity, the difference of mechanical pressure probed by each membrane must take its root in their dynamics. The importance of dynamics in determining quantities, which at equilibrium are computed from static only, is we believe a hallmark of active matter, on which future investigations shall concentrate.

{\it --- Acknowledgment ---}
We thank Michael Schindler for inspiring discussions and helpful suggestions. G. Briand is supported by Ecole Doctorale ED 564 "Physique en Ile de France". R. Ledesma-Alsonso was appointed by Ecole Sup�rieure de Physique Chimie Industrielle Paris. 
\vspace{-2mm}
\bibliographystyle{unsrt}
\bibliography{/Users/olivierdauchot/Documents/_Science/Biblio/Active,/Users/olivierdauchot/Documents/_Science/Biblio/Jamming,/Users/olivierdauchot/Documents/_Science/Biblio/Glasses,/Users/olivierdauchot/Documents/_Science/Biblio/PhyStat,./poly.bib}



\end{document}